\documentclass[runningheads,anonymous=false]{llncs}
\usepackage{graphicx}
\usepackage{xcolor}
\usepackage{amsmath}
\usepackage{mathtools}
\usepackage{hyperref}
\usepackage{amssymb}
\usepackage{pdfpages}

\newcommand{\blue}[1]{\textcolor{black}{#1}}

\newcommand{\repeatthanks}{\textsuperscript{\thefootnote}}
\DeclarePairedDelimiter\ceil{\lceil}{\rceil}

\begin{document}
\title{LeiBi@COLIEE 2022: Aggregating Tuned Lexical Models with a Cluster-driven BERT-based Model for Case Law Retrieval}
\author{Arian Askari\thanks{Both authors contributed equally to this research.}\inst{1} \and
Georgios Peikos\repeatthanks\inst{2} \and
Gabriella Pasi\inst{2} \and
Suzan Verberne\inst{1}
}
\authorrunning{A. Askari, G. Peikos, G. Pasi, S. Verberne}
\titlerunning{LeiBi@COLIEE 2022}
\institute{Leiden Institute of Advanced Computer Science, Leiden University \email{\{a.askari,s.verberne\}@liacs.leidenuniv.nl} \and Department of Informatics, Systems and Communication, University of Milano-Bicocca \email{\{g.peikos,gabriella.pasi\}@unimib.it}}
\maketitle 
\begin{abstract}
This paper summarizes our approaches submitted to the case law retrieval task in the Competition on Legal Information Extraction/Entailment (COLIEE) 2022. 
Our methodology consists of four steps; in detail, given a legal case as a query, we reformulate it by extracting various meaningful sentences or n-grams. 
Then, we utilize the pre-processed query case to retrieve an initial set of possible relevant legal cases, which we further re-rank. 
Lastly, we aggregate the relevance scores obtained by the first stage and the re-ranking models to improve retrieval effectiveness.
In each step of our methodology, we explore various well-known and novel methods. In particular, to reformulate the query cases aiming to make them shorter, we extract unigrams using three different statistical methods: KLI, PLM, IDF-r, as well as models that leverage embeddings (e.g., KeyBERT). Moreover, we investigate if automatic summarization using Longformer-Encoder-Decoder (LED) can produce an effective query representation for this retrieval task. 
Furthermore, we propose a novel re-ranking cluster-driven approach, which leverages Sentence-BERT \cite{reimers2019sentence} models that are pre-tuned on large amounts of data for embedding sentences from query and candidate documents.
Finally, we employ a linear aggregation method to combine the relevance scores obtained by traditional IR models and neural-based models, aiming to incorporate the semantic understanding of neural models and the statistically measured topical relevance. 
We show that aggregating these relevance scores can improve the overall retrieval effectiveness.
\end{abstract}

\section{Introduction}    
COLIEE is a workshop held every year since 2014 as a series of evaluation competitions related to case law. It divides four tasks into two groups: retrieval and entailment. For our participation, we have focused on Task 1, i.e., legal case law retrieval.
\par
Finding supporting precedents for a new case is critical for lawyers to fulfil their responsibilities to the court in countries with common law systems. However, due to the large number of digital legal records --- in 2021, the number of filings in the United States district courts for total cases and criminal defendants was $526,477$\footnote{\href{https://www.uscourts.gov/statistics-reports/federal-judicial-caseload-statistics-2021}{https://www.uscourts.gov/statistics-reports}} --- it requires time for legal professionals to scan specific cases and retrieve the relevant sections manually. According to studies, attorneys spend about 15 hours per week looking for case law \cite{lastres2015rebooting}.
\par
To cover this vast amount of information requests, the use of information retrieval technologies tailored to the legal field is necessary.  
Lawyers expect their search algorithms to identify all relevant cases. At the same time, in practice, they would often only analyze up to $50$ retrieved results, necessitating a precision-oriented retrieval approach~\cite{Geist}.
To fit the requirements of this search task, we leverage traditional lexical based IR models, which we optimize by tuning their parameters. By doing so, we manage to increase their precision, as we showed in prior work \cite{askari2021combining}. To further enhance the retrieval precision at high ranks, we experiment with several neural-based re-ranking models that measure the semantic similarity between the query and the top-50 retrieved documents.
\par
Specifically, in this work, we address the following questions regarding the case law retrieval task:\\
\textbf{1.} What is the retrieval effectiveness that can be obtained when various query reformulation methods are used along with traditional IR models for legal case retrieval?\label{intro:q1}\\
\noindent\textbf{2.} Can the retrieval effectiveness be improved when a re-ranking approach based on Sentence-BERT models that have pre-tuned on billions of data is employed on top of a traditional IR model?\label{intro:q2}\\
\noindent\textbf{3.} Can the aggregation of the relevance scores obtained by a traditional IR model and a neural re-ranker lead to greater retrieval precision?\label{intro:q2}
\par
We first employ various collection dependent and independent query reformulation approaches. Then, using the obtained query formats, we extensively evaluate the retrieval effectiveness of BM25, a Divergence from Randomness model, and a statistical language model. Moreover, we tune the models' associated parameters to show the significant impact of tuning these models in the studied retrieval task. In addition, we investigate the effectiveness of Vanilla BERT as a neural-based re-ranker. Finally, we aggregate the two relevance scores of best lexical model and the proposed cluster-driven re-ranker to examine whether this can further improve retrieval effectiveness. 
\section{Task description}
For a legal professional, the case law process consists of reading a new unseen case $Q_{d}$ and then, given a collection of previous cases, choosing supporting cases $S1, S2,..., Sn$ (that are called `noticed cases') to strengthen the decision for $Q_{d}$.
\paragraph{Challenges.}
\label{sec:chal}
Case law retrieval is a form of Query-by-document (QBD) retrieval, a task in which the user enters a text document -- instead of a few keywords –- as a query, and the Information Retrieval system finds relevant documents from a text corpus \cite{yang2009query,yang2018retrieval}. Transformer-based architectures \cite{vaswani2017attention}, such as BERT-based ranking models \cite{nogueira2019passage,abolghasemi2022improving} have yielded improvements in many IR tasks. However, the time and memory complexity of the self-attention mechanism in these architectures is $O (L^2)$ over a sequence of length L \cite{beltagy2020longformer,zaheer2020big}. That causes challenges in QBD tasks where we have long queries and documents. For instance, the average length of queries and documents in the legal case law retrieval task of COLIEE 2022 is more than 4000 words. Moreover, variants of Transformers that aim to cover long sequences such as Longformer \cite{beltagy2020longformer} have not shown high effectiveness which could be due to the their sparsified attention mechanism \cite{sekulic2020longformer} or the limited number of training instances for professional search tasks \cite{askari2021combining}. In this paper, we propose a cluster-driven BERT-based methodology that can consider the whole length of the query, while simultaneously overcoming the complexity imposed by length of text. Our novel re-ranking approach further improves the system's effectiveness using a weighting mechanism that combines the BM25 score obtained from the initial retrieval with the scores from the neural models.
\section{Methodology}
\label{chap:methods}
Our methodology consists of four steps, each to tackle case law retrieval: (1) Query Case Reformulation: to make queries shorter and in keyword format for lexical models as previous studies show it improves the retrieval effectiveness \cite{askari2021combining,locke2017automatic} (2) Lexical Ranker: for retrieving first stage results (3) Neural Re-ranker: for re-ranking top-$k$ candidate documents considering semantic besides of exact keyword matching (4) Relevance score aggregation: to combine \{statistical (2) and semantic (3)\}-based models together.
Question 1, 2 (as well as challenge mentioned in Section \ref{sec:chal}), and 3 that are mentioned in introduction addressed by step (1,2), (2), and (4) respectively.
\subsection{Query Case Reformulation}
In the literature, several approaches for reformulating a verbose query into either a keyword like representation or a summary have been proposed \cite{askari2021combining}. We experiment with two different approaches to create shorter query cases: (1) term extraction and (2) abstractive summarization. We do not experiment with named entity recognition and noun phrase detection methods as they have shown lower effectiveness in comparison to KLI previously on COLIEE in previous work \cite{askari2021combining}.

\noindent\textbf{Term extraction. } We experiment with three lexical-based similar to Locke et al. \cite{locke2017automatic} and one neural-based approaches for term extraction: (1) Kullback-Leibler divergence (KLI), (2) parsimonious language model (PLM) \cite{hiemstra2004parsimonious}, (3) IDF-r \cite{koopman2017generating}, (4) KeyBERT~\cite{grootendorst2020keybert}.

\paragraph{{KLI.}} We used Kullback-Leibler divergence for Informativeness (KLI) \cite{verberne2016evaluation}. The KLI score was computed for each term t in a query case document,$Q_{d}$ similar to \cite{askari2021combining}:
\begin{equation}
    KLI(t)=P(t|Q_{d})\times \log \frac{P(t|Q_{d})}{P(t|C)}
\end{equation}
where $P(t|D)$ is the probability of $t$ in the query document $D$ and $P(t|C)$ is the probability of $t$ in a background language model. We use all candidate documents as the background collection to compute $P(t|C)$.

\paragraph{{IDF-r. }} The IDF-r method selects the $\ceil{\frac{|Q_{d}|}{r}}$ terms in $Q_{d}$ \blue{with the highest Inverse Document Frequency (IDF-$r$) score where $r$ refers to the proportion of selected terms.}

\paragraph{{PLM. }} For the PLM method, we used our collection, $P(t|C)$, as the background language model and the information object, $Q_{d}$, as the foreground language model. The expectation maximization algorithm was used to estimate probabilities, with the following steps:
\begin{equation}
    E - step: e_t = tf(t,Q_{d}).\frac{\lambda .  P(t|Q_{d})}{(1-\lambda) . P(t|C) + \lambda P(t|Q_{d})}
\end{equation}
\begin{equation}
    M - step: P(t|Q_{d}) = \frac{e_t}{\sum_{t' \in Q_{d}}e_{t'}}
\end{equation}
\par
Where $\lambda \in [0, 1]$ is a smoothing parameter that controls the influence of statistics from the collection $(C)$ over the statistics from the information object $(D)$.

\paragraph{KeyBERT.}The aforementioned methods rely on document and collection statistics to extract informative terms. Recent advances allow the employment of word embeddings for term extraction. These methods can capture the semantic relationship between a document's terms and extract those that better represent its content. For our experiments, we have employed KeyBERT~\cite{grootendorst2020keybert}, which is a term extraction technique that leverages BERT-based pre-trained models. 
\par
The main idea behind KeyBERT is that those terms that have a vector representation similar to the document's vector representation, can be considered the document's most representative terms. However, legal documents can be lengthy. Moreover, a document's topic may shift from one paragraph to another. These are two characteristics of the legal domain which we had considered when we applied KeyBERT to extract representative terms from the query case. 
\par
Specifically, given a query case, we split it up into its paragraphs. Then, given each paragraph of a query case, we employ KeyBERT to create a list of candidate n-grams (bag-of-words). Having done that, KeyBERT produces an embedding representation for the whole paragraph and an embedding representation for each of its candidate n-grams. To identify the representative terms, it measures a pairwise cosine similarity score between each n-gram and the obtained paragraph embedding vector. Regarding the paragraph embedding vector, if the number of paragraph tokens exceeds the transformer's token limit, the paragraph's embedding vector is computed using mean pooling on the individual paragraph embedding vectors. However, as reported in \cite{grootendorst2020keybert}, the extracted terms are often similar to each other. To overcome this issue, KeyBERT applies an extra step to diversify the extracted terms that relies either on the Maximal Marginal Relevance (MMR) formula~\cite{DBLP:conf/conll/Bennani-SmiresM18} or a simpler formula that extracts the least similar terms based on their pairwise cosine similarity. 
\par
We employ KeyBERT following the above-mentioned steps and tuning all the required parameters using a training set. Also, we experiment with both domain-specific and non-domain-specific pre-trained models to obtain the word and paragraph embeddings. Further details related to its parameterization are reported in Section~\ref{sec:keyberttune} while the obtained experimental results are described in Section~\ref{sec:keybertretrv}.\\
\paragraph{Abstractive Sumarization}
The current state of the art in abstractive summarization is based on Transformer models \cite{lewis2019bart,raffel2019exploring}.
As the input of pre-trained available models of these architectures is limited to 1024 tokens, \cite{beltagy2020longformer} proposed Longformer-Encoder-Decoder (LED), which is a Transformer variant that supports much longer inputs. For this competition, we fine-tune LED on caselaw summaries of COLIEE $2018$ followed by the same implementation by \cite{askari2021combining} and evaluate the effectiveness of LED (hereafter mentioned as \textit{summaryQ LED fine-tuned}) for case law retrieval using three traditional IR model.

\subsection{Ranking models} 
By experimenting with different query reformulation and summarization methods (see above), we have created several representations for the studied query cases which we have evaluated in combination to various statistical and neural IR models and neural re-rankers. Specifically, we employ BM25, and we fine-tuned its parameters to the peculiarities of the studied task (lengthy documents and queries). 
Although BM25, and statistical language modelling approaches are the most well-known word-based information retrieval models, recent works do not often tune their associated parameters. Specifically, the BM25 formula contains two parameters ($k1$, and $b$) associated with the term frequency saturation and document length normalization. In the legal domain, particularly in the task of case law retrieval, tuning them is crucial, as both the considered query cases and the documents are lengthy. 

In addition, we experiment with statistical language models and neural models to investigate the effectiveness of domain-specific and non-domain-specific language modelling. Lastly, two submissions are based on a novel re-ranker on the top fifty retrieved documents. Therefore, we wanted to evaluate if some other IR models can retrieve more relevant documents in the first positions. To this aim, we employ various models that are based on the Divergence from the Randomness framework, proposed by \cite{DBLP:journals/tois/AmatiR02}; it has been found that these models can achieve higher precision compared to BM25 for some search tasks.
\subsection{Cluster-driven method (re-ranking model)}
Transformer-based models are limited in taking into account long documents, so we split the queries and documents at the sentence-level and embed the document in sentence-level using Sentence-BERT (SBERT~\cite{reimers2019sentence}). We develop a method that exploits clusters of sentences extracted from a query and a document. The proposed method, utilizes three different pre-fine-tuned SBERT's models to obtain appropriate embeddings for each part of the proposed method. Given a query $Q_{d}$, we first identify the three most important sentences and use them as query representation, inspired by the Lead-3 model \cite{nallapati2017summarunner}. Lead-3 proposes that the first three sentences of the document are good representatives for extractive summarization. To this aim, we apply K-means (k=3) as our clustering method to find three clusters of query sentences; each cluster contains semantically similar sentences. Then, we compute the centroid of the cluster and select a sentence of cluster that is most close to its centroid as the representation of the cluster. The three representative sentences are denoted as $3S_Q{_{d}}$. Then, given a candidate document $d$, we find the most similar sentences ($3S_d$) of that document to $3S_{Q_{d}}$, by computing cosine similarity. To do that, we use a bi-encoder model that is pre-fine-tuned for computing cosine similarity between two separately embedded sentences. To compute the overall relevance score between the $Q_{d}$ and each $d$ we sum the similarity scores of $3S_q$ and $3S_d$ using a cross-encoder model that is pre-fine-tuned to compute probability of relevance between two sentences. We re-rank candidate documents based on this relevance score. We miss no part of candidate query and document into our methodology due to the sentence-level embedding and considering all of sentence embeddings into our methodology for computing relevance score. In summary, our approach consists of three steps: (1) Finding the three most important sentences in the query by clustering the embeddings of query sentences using a pre-fine-tuned bi-encoder to embed sentences and k-means for clustering; (2) Finding the most similar sentences of a candidate document according to the sentences found in (1) by computing cosine similarity (3) Computing the final relevance score by computing a sum over the probability of relevance of pair of query and document sentences found in (2).
\subsection{Relevance score aggregation}\label{sec:weighting}
In the literature, it has been found that aggregating the scores of the initial ranker and re-ranker yields improvements in terms of retrieval effectiveness\textcolor{red}{\cite{askari2021combining,althammer2021dossier,askari2022expert}}.
Our experiments also observed this behaviour as it was found that re-ranking improves several queries but hurts others. 
One of the most common methods is the linear aggregation of the two obtained relevance scores. The aggregation approach mentioned above associates an importance value to the relevance score of the ranker ($S_BM25$) and one value to the relevance score of the re-ranker ($S_{cluster-driven}$). Similar to \cite{althammer2021dossier}, the final relevance score is computed as below to find the optimal value for weighting method (hereafter mentioned as \textit{weighting}) on the validation set where $\alpha, \beta \in \mathbb{R}$:
\begin{align*}
    s_{BM25+cluster-driven}(d,Q_{d}) = \alpha \ s_{BM25}(d,Q_{d}) + \beta \ s_{cluster-driven}(d,Q_{d}).
\end{align*}

\section{Experiment design}
\label{sec:experiments}
\subsection{Collection}
\label{sec:data}
For our experiments, we use the collection provided by the organizers of COLIEE'22 \cite{goebel2022coliee}. Precisely, the collection consists of a training and a test set. The test set contains $300$ query cases associated with $1,563$ relevant documents. The training set includes $898$ query cases and $4,415$ relevance assessments. To fine-tune our models, we have split the train set into train and validation sets. The validation set in which we report our results was created by selecting the last $250$ queries out of $898$ training queries. Hereafter, when we refer to the training set, we refer to the $648$ remained queries. 
\subsection{Query pre-processing}
\subsubsection{Tuning the KLI \& PLM methods.}
Both the KLI and the PLM methods score each document term, creating a term ranking. As a result, one can identify which proportion of this ranking can better represent its content. We found the optimal term proportion based on optimizing it according to F1 score.

\subsubsection{Tuning KeyBERT.}
\label{sec:keybertretrv}
To extract essential terms for each document paragraph using KeyBERT, we use its official implementation, which is publicly available~\footnote{\url{https://github.com/MaartenGr/KeyBERT}}. Regarding its parametric setup, there are several parameters in this implementation, such as the size of the extracted n-grams, the number of extracted terms, and the pre-trained embedding model. In addition, one can choose between two different term diversification methods (Max Sum Similarity and MMR) and tune their corresponding diversification coefficient. We tuned the parameters on the validation set. 
\par
For our experiments, we utilize two pre-trained models to obtain the embedding representations, namely the \textit{all-MiniLM-L6-v2} sentence transformer~\footnote{\url{https://huggingface.co/sentence-transformers/all-MiniLM-L6-v2}} and \textit{Legal BERT} \cite{chalkidis-etal-2020-legalbert}. In addition, we experimented with both the Maximal Marginal Relevance (MMR) diversification and the Max Sum Similarity formulas and tuned their diversification coefficient value in range [.2-.8] using a step of .1. Moreover, we set the parameter related to the n-gram size so that KeyBERT produces uni-grams and bi-grams. Finally, we experimented by altering the number of the extracted term from each paragraph in the range [5-25] using a step of 5. To tune all the required parameters, we have used the train set described in Section~\ref{sec:data}. We concatenated all the obtained n-grams into one query text and used it for retrieval. As optimal parameters, we have chosen those that maximized the F1 score in the considered training set.

\subsection{Lexical rankers}
\paragraph{\textbf{{BM25}}} We ran BM25 using the default parameter values of $k1 = 1.2$ and $b = 0.75,$ and its implementation from ElasticSearch. In addition, we created \textit{BM25-opt} by \blue{tuning its hyperparameter with doing grid search over $b$ $\in$ $\{0, 0.1, 0.2, ·, 1\}$ and $k=$ $\in$ $\{0, 0.1, 0.2, ·, 3\}$} on the validation set.

\paragraph{\textbf{{Language modelling}}} We used the built-in similarity functions of Elasticsearch for the implementation of Language Modelling (LM) with two different smoothings: Dirichlet smoothing and Jelinek Mercer (JM) smoothing. We only report the results for JM smoothing since we get similar results from these two smoothing methods. We also optimised the hyperparameter value ($\lambda$) for Language Modelling with Jelinek Mercer smoothing (LM JM). \blue{We found $\lambda$ = 0.1 as the optimal value for LM JM with KLI.}

\paragraph{\textbf{Divergence from Randomness \blue{(DFR)}}} To implement the DFR models, we have used PyTerrier \cite{pyterrier2020ictir}, and we tuned their associated parameter, using the training set, aiming at optimising the P@4 measure. Specifically, these models are associated with a free parameter $c$ that controls the term frequency normalisation component\cite{DBLP:journals/tois/AmatiR02}.
We found that for several DFR models ---In\_expC2, In\_expB2, lnL2, lnB2, described in \cite{DBLP:journals/tois/AmatiR02}--- the default parameter setting (c=.1) was the optimal value.

\paragraph{\textbf{Cluster-driven method. }}
We use \textit{all-mpnet-base-v2}, \textit{msmarco-bert-base-dot-v5}, and \textit{ms-marco-MiniLM-L-12-V2} models of Sentence-BERT for clustering, computing cosine similarity and computing probability of relevance respectively

\section{Results}
This section presents the top-performing retrieval models' experimental results on the validation set. Table~\ref{tab:retresults_lex} presents the retrieval effectiveness obtained across several query representations and models, while Table~\ref{tab:rerankesults_valset} presents the results obtained by the re-ranker and the score weighting method. For every retrieval model and query representation method presented in Tables~[\ref{tab:retresults_lex},\ref{tab:rerankesults_valset}], the reported parameters were those that optimized the F1 score on the training set.

\begin{table}[t]
\caption{BM25, LM Jelinek Mercer (JM) and DFR retrieval results for the ranking of candidate documents on the validation set of COLIEE'22. We only report result of optimized DFR In\_expC2.}
\centering
\resizebox{!}{!}{%
    \begin{tabular}{l|l|c|c|c}
        Method    & Query representation & P \% & R \%  & F1 \% \\ \hline
        \hline
        BM25 (not optimized) & original text & 13.10 & 19.09 & 15.53 \\ \hline
        BM25 & original text & 16.30 & 19.34 & 17.69 \\ \hline %
        BM25 & KeyBERT (Legal BERT) & 6.50 & 10.29 & 7.96 \\ \hline %
        BM25 & KeyBERT (all-MiniLM-L6-v2) & 6.55 & 9.80 & 7.85 \\ \hline %
        BM25 & KLI (1-gram/40\% term portion) & \textbf{19.30} & \textbf{28.59} & \textbf{23.04} \\ \hline
        BM25 & PLM (1-gram/50\% term portion) & 17.20 & 25.91 & 20.67 \\ \hline
        BM25 & IDF (1-gram/90\% term portion) & 15.80 & 23.11 & 18.76 \\ \hline
        BM25 & summaryQ LED fine-tuned & 5.73  & 12.84 & 7.92 \\
        \hline \hline
        LM JM (not optimized) & original text & 12.19 & 18.54 & 14.70 \\ \hline
        LM JM  & original text & 17.10 & 25.32 & 20.41 \\ \hline %
        LM JM  & KeyBERT (Legal BERT) & 6.23 & 10.90 & 7.92 \\ \hline %
        LM JM  & KeyBERT (all-MiniLM-L6-v2) &  6.02 & 10.98 & 7.77 \\ \hline %
        LM JM  & KLI (1-gram/40\% term portion)  & 15.76 & 29.47 & 20.53 \\ \hline
        LM JM  & PLM (1-gram/50\% term portion) & 16.50 & 24.05 & 19.57 \\ \hline
        LM JM  & IDF (1-gram/90\% term portion)  & 5.87 & 12.01 & 7.88 \\ \hline
        LM JM  & summaryQ LED fine-tuned & 5.07  & 11.73 & 7.07 \\
        \hline \hline
        DFR In\_expC2 & original text & {10.80} & {15.87} & {12.85} \\ \hline
        DFR In\_expC2  & KeyBERT (Legal BERT) & 9.40 & 13.39 & 11.04 \\ \hline %
        DFR In\_expC2  & KeyBERT (all-MiniLM-L6-v2) & 11.00 & 15.60 & 12.91 \\ \hline %
        DFR In\_expC2  & KLI (1-gram/40\% term portion) & 12.70 & {19.07} & {15.24} \\ \hline
        DFR In\_expC2  & PLM (1-gram/50\% term portion) & 12.50 & 18.18 & 14.82 \\ \hline
        DFR In\_expC2  & IDF (1-gram/90\% term portion) & 9.80 & 14.29 & 11.62 \\ \hline
        DFR In\_expC2  & summaryQ LED fine-tuned & 7.60  & 10.93 & 8.96 \\\hline\hline
    \end{tabular}
}
\label{tab:retresults_lex}
\end{table}
\paragraph{\textbf{Summary of results}} The results show that the optimal retrieval effectiveness is achieved when the KLI method, with the reported parameters, is combined with the BM25 model. Therefore, this retrieval combination was selected as one of our submitted runs. Moreover, another remark is related to the fact that all of the query representations that were created using embedding based models lead to poor retrieval effectiveness. Finally, we observe that all of the employed query reformulation approaches seem robust and yield similar results across the employed retrieval models.
\paragraph{\textbf{Retrieval using the KeyBERT terms.}}
\label{sec:keyberttune}
From each paragraph of a query case, KeyBERT returns a pre-defined number of extracted n-grams that are later concatenated to create a single query representation used for retrieval. Using the train set, we have identified the optimal parameter setting. In particular, it has been found that the optimal number of extracted n-grams is $20$, the diversification coefficient is $0.6$, while it was found that the MMR term diversification formula leads to greater improvements in terms of F1 score on the training set compared to the Max Sum formula. Moreover, it has been found that these optimal parameters remained unchanged when we used the domain-specific pre-trained model to obtain the embedding representations. Therefore, we have investigated the effectiveness of the KeyBERT term extraction method, using the above mentioned parameters in combination with either a domain specific pre-trained model or a non-domain specific pre-trained model to obtain the embedding representation. The obtained experimental results on the validation set are presented in Table~\ref{tab:retresults_lex}.
\paragraph{\textbf{Retrieval using DFR models.}} By altering the basic models used to calculate the probabilities in the generic DFR formula along with the term frequency normalization, one may obtain several DFR models.
We have experimented with several variation of DFR retrieval models, using the original query text as input. Results have shown that the Inverse Expected Document Frequency model with Bernoulli after-effect and normalisation 2, namely \textit{In\_expC2}, yields the best performance. 
As a result, we used only this model in combination with the various query representation. The obtained results are presented in Table~\ref{tab:retresults_lex}.
\begin{table}[t]
\caption{Retrieval results for the ranking of candidate documents on the validation set of COLIEE'22, using cluster-driven as re-ranker and weighting as score aggregator between BM25 and cluster-driven method. BM25 optimized reported for comparison.}
\centering
\resizebox{!}{!}{%
    \begin{tabular}{l|l|l|l|l}
        Method & Query representation & P \% & R \%  & F1 \% \\ \hline
        \hline
        BM25 & KLI (1-gram/40\% term portion) & 19.30 & \textbf{28.59} & 23.0  \\ \hline
        Cluster-driven & original text (sentence-level) & 16.20 & 23.31 & 19.11 \\ \hline
        Weighting & - & \textbf{19.75} & 28.29 & \textbf{23.26} \\ \hline
    \end{tabular}
}
\label{tab:rerankesults_valset}
\end{table}
\paragraph{\textbf{Neural re-ranker}}
We re-rank the top-$50$ candidate documents retrieved using the BM25-optimized model. While, as shown in Table~\ref{tab:rerankesults_valset}, the BM25-optimized outperforms our cluster-driven method in terms of mean effectiveness, our investigation show that there are queries for which our cluster-driven method improves their performance over the BM25-optimized.
As a result, the neural re-ranker is effective when considered in a weighting aggregation method (See section~\ref{res:weighting}).
\par
It is noteworthy to mention that we experimented with Vanilla BERT following our previous works on COLIEE in \cite{askari2021combining}. However, we achieved lower results than BM25, and we found that applying the weighting mechanism on the scores obtained by Vanilla BERT and BM25 could not outperform BM25. The best F1 that we could achieve by BERT was $18.47$ on the validation set, which is lower than the results achieved by cluster driven method on the same set (F1 $19.19$). As a result, we only exploit the proposed cluster-driven method as our neural re-ranker in this paper. We submitted cluster-driven method to assess its effectiveness on test set in the COLIEE 2022 competition as the second submission.
\paragraph{\textbf{Weighting. }}\label{res:weighting}
We use the weights $[1,2,...,100]$ for $\alpha$ and $\beta$ values \blue{(See section \ref{sec:weighting})}. Optimal weights found as $36$ for cluster-driven method and $48$ for BM25. From Table~\ref{tab:rerankesults_valset} it is clear that aggregating the two relevance scores obtained by the KLI+BM25 retrieval pipeline and the cluster-based re-ranker, further improves the retrieval effectiveness on the validation set. Therefore, that approach was our third submission.

\section{Discussion}
\begin{figure*}[t]
    \centering
    \includegraphics[width=0.80\textwidth]{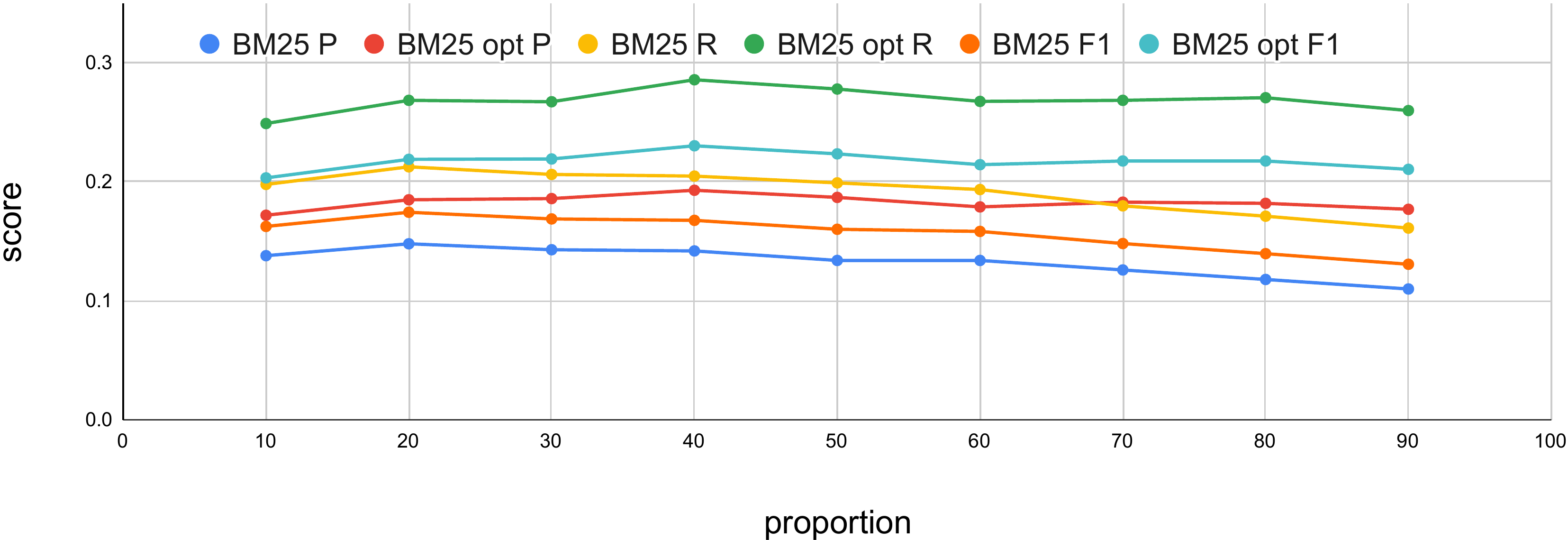}
    \includegraphics[width=0.80\textwidth]{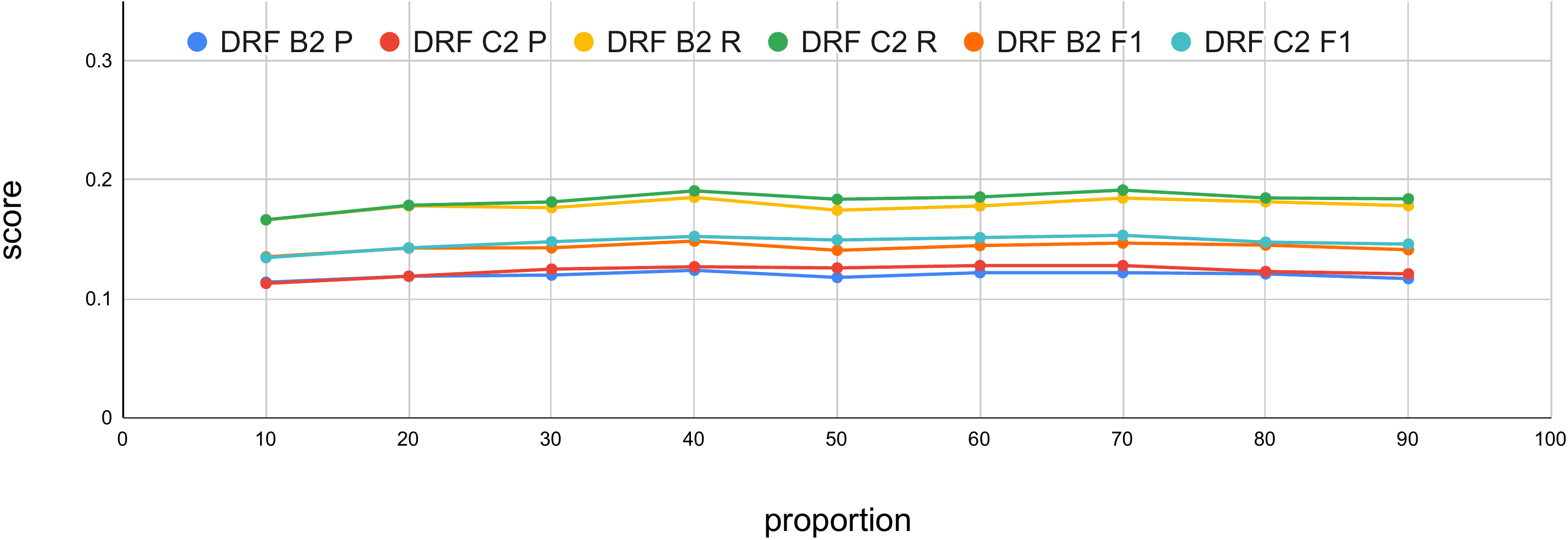}
    \caption{Retrieval results for BM25 and BM25 opt with default and optimized parameters and DFR (B2 and C2) on the validation set of COLIEE'22 with using different proportion of terms scored by KLI method as a shorter query. P and R refers to Precision and Recall respectively.}
    \label{fig:kli_proporiton_validation}
\end{figure*}

\paragraph{\textbf{Proportion of KLI terms}}
Table \ref{tab:retresults_lex} show the effect of query representation methods on lexical retrieval models. Statistical term extraction methods (KLI, PLM, IDF) shows higher effectiveness in compare to KeyBERT and Longformer-Encoder-Decoder (LED) and KLI shows the most effectiveness within all three rankers (BM25, LM JM, DFR ln\_expC2). BM25 achieve highest effectiveness using query terms that are extracted by KLI.
We analyze the effect of using different proportion of terms scored by our best term extraction method KLI on BM25 (our best lexical ranker) and DFR to assess the sensitivity of BM25 on using different proportions of terms and compare BM25's sensitivity with DFR as another lexical ranker.
The figures \ref{fig:kli_proporiton_validation} show BM25 has higher sensitivity on using different proportion of terms in comarparison to DFR ln\_expC2. We interpret this sensitivity as a strength of BM25 in this case because using top-$40$\% of query terms as the query improve the effectiveness of BM25 more than DFR while both rankers receive highest effectiveness with using same proportion of terms extracted by KLI. On both rankers, increasing the proportion of terms till $40$\% improves the recall and consequently F1 while the precision is almost consistent. This can be related to the fact that enriching the query with more approperiate words can increase the chance of retrieving relevant documents that are mathced with that added words in lexical rankers.

\begin{figure}[!htb]
   \begin{minipage}{0.48\textwidth}
     \centering
     \includegraphics[width=1.0\linewidth]{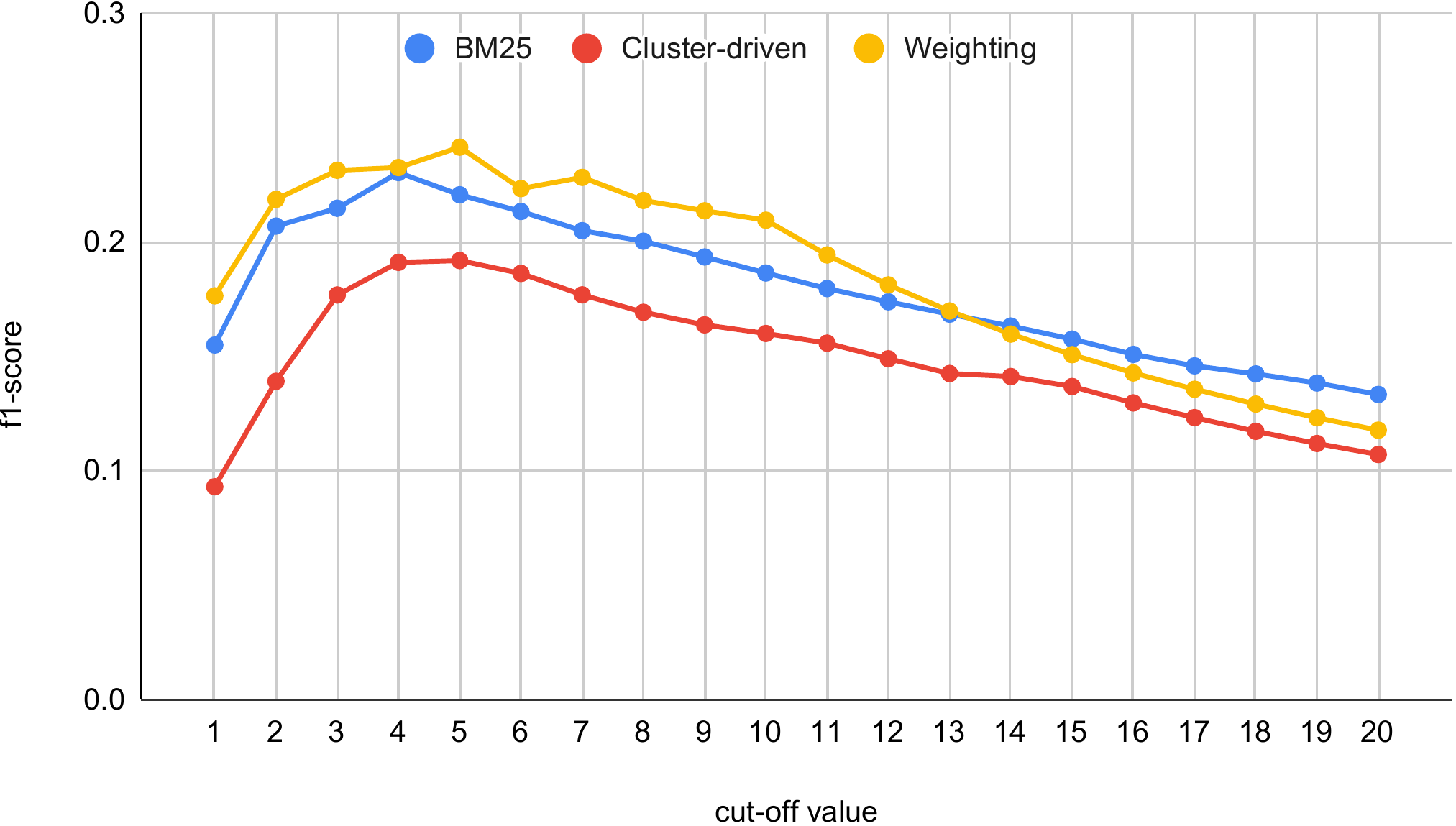}
     \caption{F1-score for Task 1 on the \textbf{validation} set for different re-ranking depth (cut-off value for first-stage ranker)}\label{fig:evaltask1}
   \end{minipage}\hfill
   \begin{minipage}{0.48\textwidth}
     \centering
     \includegraphics[width=1.0\linewidth]{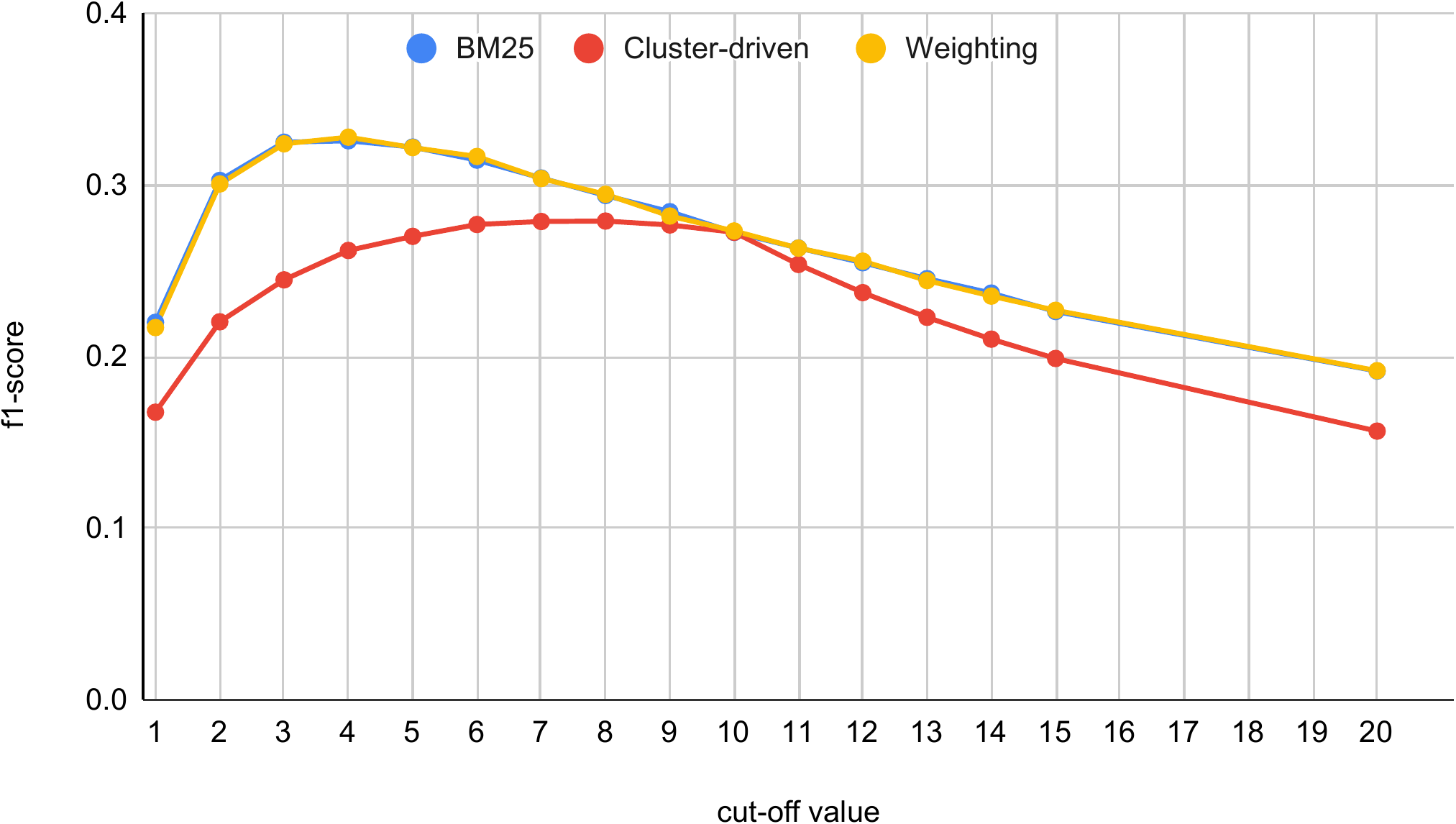}
     \caption{F1-score for Task 1 on the \textbf{test} set for different re-ranking depth (cut-off value for first-stage ranker)}\label{fig:testtask1}
   \end{minipage}
\end{figure}

\paragraph{\textbf{Effect of the cut-off value.}}
As the task is to retrieve the relevant cases to a given query $q$, we consider the top-k-ranked documents $d$ in the ranked list as relevant and denote $k$ as cut-off value.  We evaluate the best cut-off value $k$ depending on the F1 score of the validation set using pytrec\_eval\footnote{https://github.com/cvangysel/pytrec\_eval}. The results are shown in Figures~\ref{fig:evaltask1} and~\ref{fig:testtask1} show the effect of cut-of value on F1 score. We found cut-off $4$ as optimal value for BM25 and cluster-driven method and $5$ for weighting method. In these figures, BM25 refers to the BM25 optimized.
\begin{table}[t]
\caption{Retrieval results for the ranking of candidate documents on the test set of COLIEE'22, using our top-performing approaches.}
\centering
\resizebox{!}{!}{%
    \begin{tabular}{l|l|l|l|l}
        Method & Extractor/Sumarizer & P \% & R \%  & F1 \% \\ \hline
        \hline
        BM25 optimized & KLI (1-gram) & \textbf{29.92} & 35.75 & 32.57 \\ \hline
        Cluster driven & original text (sentence-level) & 23.92 & 28.92 & 26.18 \\ \hline         
        Weighting & - & \textbf{29.92} & \textbf{36.26} & \textbf{32.78} \\ \hline
    \end{tabular}
}
\label{tab:submitted_runs}
\end{table}

\paragraph{\textbf{Submitted Experiments}} 
We submitted three run files for task 1: (1) BM25-optimized with KLI as our term extraction method; (2) the cluster-driven method as a re-ranker using BM25-optimized with KLI as initial ranker; (3) a weighting model that aggregates the output scores of (1) and (2) using linear aggregation. Table~\ref{tab:submitted_runs} shows that the weighting mechanism could improve BM25 optimization by utilizing both lexical (obtained by BM25) and neural-based matching (obtained by cluster-driven) scores.
\section{Conclusion and future work}
Our participation at COLIEE 2022 in Task 1 allowed exploring the effect of term extraction methods on lexical and neural models in case law retrieval. We identify that the presence of long documents creates significant issues both for neural re-ranking strategies and statistical models. Therefore we propose, compare and evaluate a cluster-driven BERT-based re-ranker that considers the whole length of the query and candidate documents. The model could outperform Vanilla BERT when used with BM25 as the initial ranker. 
\par
We show that optimized BM25 is the best lexical matching model among LM and DFR models. KLI is the best term extraction method compared to PLM, IDF-r, and KeyBERT. We find that the combination of lexical and neural models improves the overall effectiveness of the ranking. In the future, we plan to investigate further the effectiveness of our novel cluster-driven model, by considering the importance of each cluster, in the estimation of the overall relevance score.

\section{Acknowledgement}
This project has received funding from the European Union’s Horizon 2020 research and innovation programme under the Marie Skłodowska-Curie grant agreement No 860721  – DoSSIER (H2020-EU.1.3.1.).

\bibliographystyle{splncs04}
\bibliography{ref.bib}
\end{document}